\documentclass[12pt,reqno]{amsart}
\usepackage{amsmath,amssymb,amsfonts,amsthm}
\usepackage[mathscr]{eucal}
\usepackage[all]{xy}
\usepackage{hyperref}
\usepackage{setspace}
\allowdisplaybreaks

\author{D.S.~Kaparulin, S.L.~Lyakhovich}

\address{Department of Quantum Field Theory, Tomsk State University, Lenin ave. 36, Tomsk 634050, Russia.}

\title{A note on unfree gauge symmetry}
\begin{document}

 \maketitle

\begin{abstract}
We study the general structure of field theories with the unfree
gauge symmetry where the gauge parameters are restricted by
differential equations. The examples of unfree gauge symmetries
include volume preserving diffeomorphisms in the unimodular gravity
and various higher spin field theories with transverse gauge
symmetries. All the known examples of the models with unfree gauge
symmetry share one common feature. They admit local quantities which
vanish on shell, though they are not linear combinations of
Lagrangian equations and their derivatives. We term these quantities
as mass shell completion functions.

\noindent In the case of usual gauge symmetry with unconstrained
gauge parameters, the irreducible gauge algebra involves the two
basic constituents: the action functional and gauge symmetry
generators. For the case of unfree gauge symmetry, we identify two
more basic constituents: operators of gauge parameter constraints
and completion functions. These two extra constituents are involved
in the algebra of unfree gauge symmetry on equal footing with action
and gauge symmetry generators.

\noindent Proceeding from the algebra, we adjust the Faddeev-Popov
(FP) path integral quantization scheme to the case of unfree gauge
symmetry. The modified FP action involves the operators of the
constraints imposed on the gauge  parameters, while the
corresponding BRST transformation involves the completion functions.
The BRST symmetry ensures gauge independence of the path integral.
We provide two examples which admit the alternative unconstrained
parametrization of gauge symmetry and demonstrate that they lead to
the equivalent FP path integral.
\end{abstract}

\section{Introduction}
In this article we study the gauge symmetry with unfree
transformation parameters. The usual definition of gauge symmetry
implies that the transformation parameters are arbitrary functions
of space-time coordinates.  The set of the gauge symmetry generators
can be over-complete that results in gauge symmetries of symmetries.
This is known as a reducible gauge symmetry. At the level of
corresponding Batalin-Vilkovisky (BV) formalism
\cite{BV1},\cite{BV2} and even when the Faddeev-Popov (FP) method
works well, the symmetry for symmetry requires to introduce ghosts
for ghosts. The unfree gauge symmetry is a different phenomenon.  It
is the symmetry with the transformation parameters subject to the
differential equations. Some articles term these equations as
constraints imposed on the gauge transformation parameters
\cite{Francia:2013sca}, \cite{Francia:2016weg}. We can also mention
the phenomenon of so-called semi-local symmetries \cite{Sorokin},
\cite{Beckaert}, \cite{Bandos} which is somewhat similar to the
unfree gauge symmetry because the gauge parameters have to obey the
differential equations in both cases. The difference is that the
solutions of the equations on the unfree parameters involve the
arbitrary functions of $d$ coordinates in $d$ dimensional space,
while in the semi-local case the arbitrary functions depend on $d-1$
coordinates or less.

One of the best known examples of the field theory with unfree gauge
symmetry is the unimodular gravity \cite{Unruh:1988in},
\cite{Buchmuller}, \cite{Henneaux}, \cite{Ellis1}, \cite{Ellis2},
where the gauge symmetry reduces to the volume-preserving
diffeomorphisms, so called $T$-diff's. The $T$-diff is the usual
diffeomorphism with the unfree transformation parameter which is
supposed to be a transverse vector field. The gauge parameter is
constrained by the differential equation: the divergence should
vanish of the vector. Also notice that the $T$-diff's form a
subalgebra in the full algebra of diffeomorphisms. In some cases,
this subalgebra can be treated differently from the entire
diffeomorphism in the Hamiltonian
BFV-BRST\footnote{Batalin-Fradkin-Vilkovisky (BFV);
Becchi-Ruet-Stora-Tyutin (BRST)} quantization of gravity
\cite{Kamenshchik:1996hb} even if the unimodularity condition is not
imposed. While the unimodular gravity is equivalent to General
Relativity (GR) at classical level, at least in most of physical
problems, some distinctions are possible between the GR and
modifications of unimodular gravity, see \cite{Kamenshchik},
\cite{Barvinsky}. There are also some subtleties about the
equivalence at quantum level, for discussion see
\cite{Padilla:2014yea}, \cite{Percacci1}, \cite{Percacci2} and
references therein. Even though the unimodular gravity can be
considered as a reformulation of the GR, these two theories differ
by the gauge symmetry, so the general structure of unfree gauge
symmetry remains the problem of interest for this model. Various
advantages and disadvantages are known of different forms of gauge
symmetry of locally equivalent formulations of gravity
\cite{Gielen:2018pvk}.

Besides the $T$-diff's in gravity, various models with  unfree gauge
symmetry  are known among higher spin field theories, see
\cite{Skvortsov:2007kz}, \cite{Campoleoni:2012th} and the references
therein. The model of irreducible spin 2 traceless field is known
\cite{Alvarez:2006uu}, \cite{Blas:2007pp} with the vector gauge
parameter restricted by transversality equation. This model
corresponds to the linearized unimodular gravity. The work
\cite{Skvortsov:2007kz} can be viewed as an extension of the model
\cite{Alvarez:2006uu} to the higher spin fields. The unfree gauge
symmetry of the model of \cite{Skvortsov:2007kz} can be viewed as a
higher spin extension of the linearized $T$-diff's. The Maxwell-like
models of ref. \cite{Campoleoni:2012th} describe higher spin fields
in terms of a tracefull tensors. These models also have the gauge
symmetry with transverse parameters.

In the article \cite{Francia:2013sca}, the issue of the unfree gauge
theory is considered for general linear local field theory. It is
found that any linear field theory always admits unconstrained
parametrization of gauge symmetry, with the gauge parameters being
arbitrary functions unrestricted by any equation. The unconstrained
form of the gauge symmetry is reducible, in general, and the
reducibility order is always finite, being bounded by the space-time
dimension. This means, the generators of unfree gauge symmetry in
linear theory can be always replaced by equivalent (maybe
over-complete) generating set with unconstrained parameters. In the
article \cite{Francia:2013sca}, the reducible unconstrained gauge
symmetry generating sets are explicitly identified for various
higher spin theories where the gauge symmetry has been previously
known only in the constrained form.

As far as every theory admits unconstrained (possibly reducible)
parametrization of gauge symmetry at least at linearized level, it
may seem unnecessary to study the unfree gauge symmetry at all. We
can mention at least three reasons why it can be the issue of
interest.

First, the unfree gauge symmetry parametrization is equivalent, in
principle, to the unconstrained but reducible parametrization. From
technical viewpoint, and geometrically, these two parameterizations
can be quite different,  so each one may have its own advantages and
disadvantages. For example, in the model of irreducible higher spin
field described by traceless tensors proposed in ref.
\cite{Skvortsov:2007kz}, the unfree gauge transformations involve
only first order derivatives of the gauge parameters, while the
parameters are symmetric traceless tensors. These parameters are
constrained by the first order differential equations. The
unconstrained reducible transformations are found for this model in
the work \cite{Francia:2013sca}. The gauge parameters of
unconstrained symmetry are the tensors with a two-row Young
tableaux, and the transformations involve the higher order
derivatives of the parameters. The sequence of symmetry for symmetry
transformations of the free gauge parameters involve the tensor
parameters with various Young tableaux, depending on spin.
Obviously, these two algebraic structures are essentially different
while they describe the gauge symmetry of the same field theory.
Each of them can provide different insight into the dynamics once
the issues are considered of including consistent interactions or
quantization.

Second, the unconstrained gauge symmetry parametrization, with
possible reducibility, is proven to always exist
\cite{Francia:2013sca} only in the linear theory. So, if the theory
has the gauge symmetry with constrained gauge parameters at
interacting level, the unconstrained equivalent can be non-existent
at all for the nonlinear model, and vice versa. The examples are
known of the field theories which admit different unconstrained
parametrization of gauge symmetry, including reducible and
irreducible generating sets at linear level \cite{Francia:2013sca},
\cite{Lyakhovich:2014soa}. Also the example is known of the model
where the reducible generating set of the linear theory admits
consistent inclusion of interaction, while the irreducible
generating set obstructs the same interaction
\cite{Lyakhovich:2014soa}. So, the reducible and irreducible
parametrization of gauge symmetry can be inequivalent with respect
to deformations, while both generating sets can correspond to the
same model at linear level.

The third reason is that the clarification of general unfree gauge
symmetry structure can be considered as a matter of principle for
the general theory of gauge systems. The general structure of gauge
algebra is well known in the theories with unconstrained gauge
parameters both with irreducible and reducible generating sets of
gauge symmetry transformations. In particular, all the structure
relations are known for the gauge generators, the generators of
symmetry for symmetry, and the structure functions involved in the
off-shell disclosure of the symmetry. At the level of the BRST
field-anti-field formalism, all the structure relations are
generated by the BV master equation. As a standard reference to the
gauge algebra structure, we mention the textbook
\cite{Henneaux:1992ig}. The general gauge algebra also has well
developed formalism for not necessarily Lagrangian gauge field
theories \cite{Lyakhovich:2004xd}, \cite{Kazinski:2005eb}. One of
the important distinctions of the gauge algebra in the
non-Lagrangian case is that the  gauge symmetries are not
necessarily paired with gauge identities. In the Lagrangian case the
same gauge generators are involved in the gauge symmetry
transformations and in the gauge identities between the equations of
motion. In non-Lagrangian case, the generators of gauge symmetries
and gauge identities can be different in general, so the gauge
algebra involves more generating elements. The non-Lagrangian
extension of the master equation \cite{Kazinski:2005eb},
\cite{Lyakhovich:2005mk} generates  a more reach gauge algebra that
involves more structures comparing to the Lagrangian case. Any
deformation of the theory, be it inclusion of interaction or
quantization, should consistently deform all the structure relations
of the gauge algebra. Somewhat similar phenomenon we shall observe
in the theory of the systems with unfree gauge symmetry. There are
extra generating elements in the gauge algebra besides the action
and the generators of gauge symmetry transformations. These extra
elements contribute to the gauge identities. In this sense, the
unfree gauge symmetry is similar to the gauge algebra of
non-Lagrangian systems. Any deformation of the theory with unfree
gauge symmetry, either by inclusion of interaction or by
quantization, should be consistent with deformation of corresponding
gauge algebra structures. That is why it seems important to clarify
the structure of the gauge algebra with unfree parameters.

In this work we consider three aspects of the problem of unfree
gauge symmetry. At first we identify the basic generating elements
of the unfree gauge symmetry algebra. Besides the action and the
generators of gauge symmetry, there are two extra constituents
having no analogues in the usual unconstrained gauge algebra. The
first extra element is the operator of  constraint imposed onto the
gauge parameter. The second one is the structure we term the
completion function. It is explained in the next section. These two
elements define both the Noether identities and unfree gauge
symmetry of the action. The gauge symmetry transformations and gauge
identities have to satisfy the compatibility conditions that can be
viewed as the higher structure relations of the gauge algebra in the
case of unfree gauge parameters. These relations are distinct from
the analogues in the case of the gauge symmetry with unconstrained
gauge parameters. The consistent deformation of any model with
unfree gauge parameters should be compatible with these structure
relations.

In the second instance, we extend the FP path integral quantization
scheme to the theories with unfree gauge transformation parameters.
The key idea of the extension is that the ghosts for the unfree
gauge symmetry should be unfree themselves. To put it different, the
ghost should be constrained by the same equations as imposed on the
gauge parameters. Of course, the FP recipe remains applicable as far
as there are no off-shell disclosure of the gauge algebra. For the
general case of open unfree algebra, the extension of the BV master
equation has to be worked out. This problem will be addressed
elsewhere, though in the concluding section we mention some clues to
the issue. We also consider the BRST symmetry of the FP action. In
the case of unfree symmetry, the BRST transformation squares to zero
in general not identically, but modulo constraints imposed on the
ghosts.

Third, in Section 4, we consider the specific models with unfree
gauge symmetry to exemplify the general formalism and to verify the
general conclusions by alternative methods admitted by specific
models. In this section, we also exemplify the way in which the BRST
symmetry can distinguish the physical vertices from the nonphysical
ones in the models with unfree gauge parameters.

\section{Completion of Lagrangian equations and unfree gauge symmetry}
We begin this section by noticing the phenomenon which is common for
all the known examples of the field theories with unfree gauge
transformation parameters. In all these models, one can find the
on-shell vanishing local quantities such that do not reduce to
differential consequences of Lagrangian equations. These quantities
vanish by virtue of Lagrangian equations \emph{and} the boundary
conditions imposed on the fields.

For example, for the spin 2 field theory with the traceless tensor
$h_{\mu\nu}$ proposed in the article \cite{Alvarez:2006uu}, the
Lagrangian equations have the differential consequence,
\begin{equation}\label{ddd}
    \partial_\mu \tau \approx 0, \qquad  \tau=\partial_\mu\partial_\nu h^{\mu\nu} \, , \quad h^\mu_\mu\equiv 0\, .
\end{equation}
Throughout the paper, the on-shell equality is denoted as $\approx$.
Maxwell-like field theories of higher spins proposed in
\cite{Campoleoni:2012th} have the same consequence while the tensor
$h_{\mu\nu}$ is tracefull. This means, the quantity $\tau$ should be
constant on-shell. Given zero boundary conditions for the fields at
the space infinity, the constant should be zero\footnote{Notice that
the Cauchy data are not included into the boundary conditions. The
boundary conditions define the class of admissible fields, while the
Cauchy data concern the initial state of the fields. This can be
rephrased in slightly different wording: the boundary conditions
define the configuration space of the fields, while Cauchy data
define the initial field configuration and velocity.}. With the
usual boundary conditions, we have the on-shell vanishing local
quantity
\begin{equation}\label{tau}
    \tau \approx 0\, ,
\end{equation}
while $\tau$ does not reduce to linear combination of Lagrangian
equations and their derivatives. Also notice that the equation
(\ref{tau}) does not restrict solutions of the Lagrangian equations,
given the boundary conditions. For the Maxwell-like equations of
higher spin fields, this fact is emphasized in the article
\cite{Francia:2016weg}.

Given these observations made in the specific models with unfree gauge symmetry, below we
consider the general field theory where the system of Lagrangian
equations is incomplete in certain sense. We mean that the local
on-shell vanishing quantities $\tau_a$ exist such that
\begin{equation}\label{tau-a}
    \tau_a (\phi) \approx 0\, , \qquad \tau_a(\phi)\neq K_a^i\partial_i
    S(\phi) \, .
\end{equation}
Hereinafter, we use the DeWitt condensed notation\footnote{ All the
indices are condensed, in the sense that they include the space-time
coordinates. In particular, the fields $\phi^i$ are labeled by the
index $i$ which includes all the discrete indices, and the
space-time point $x$. For example, for the vector field $A_\mu(x)$,
the condensed index would include $\mu$ and $x$. Summation in the
condensed index includes integration over space-time. The
derivatives in $\phi^i$ are understood as variational, so
$\partial_iS(\phi)$ is the left hand side of Lagrange equations. },
in particular the indices $a$ and $i$ are condensed. By
$K_a^i(\phi)$ we mean the (rectangular) matrix of the local
differential operator. To put it slightly different, the Lagrange
equations $\partial_iS(\phi)\approx 0$ are incomplete in the sense
that the on shell  vanishing local quantity does not necessarily
reduce to the linear combination with local coefficients of the left
hand sides of Lagrangian equations.

We term the local quantities $\tau_a(\phi)$ as the generating set of \emph{completion
functions} if any on-shell vanishing local
quantity can be spanned in the left hand sides of Lagrange equations
and $\tau$'s:
\begin{equation}\label{Completion}
O(\phi)\approx 0\quad \Leftrightarrow\quad O(\phi)=V^i(\phi)
\partial_i S(\phi) + V^a(\phi) \tau_a(\phi) \, .
\end{equation}
The coefficients $V^i, V^a$ stand for the local differential
operators.

The usual definitions of the gauge field theory, see
\cite{Henneaux:1992ig}, assume that the Lagrange equations are
complete, in the sense that any on-shell vanishing local quantity is
a linear combination of the lhs of the equations and their
derivatives. As we learn from the examples, this assumption is not
true for the models with unfree gauge parameters.
The key observation is that the generating set for the on-shell vanishing
local quantities includes both Lagrangian equations and the
completion functions (\ref{tau-a}).

Notice that the generating set of completion functions is defined
modulo linear combinations. The generating sets of completion
functions are considered equivalent if they differ by the lhs of
Lagrangian equations,
\begin{equation}\label{tplusds}
    \tau'_a(\phi)\sim\tau_a(\phi)\, , \quad \tau'_a(\phi) =\tau_a(\phi)+\theta_a^i(\phi) \partial_iS(\phi) \, .
\end{equation}

The set of completion functions (\ref{tau-a}) can be over-complete,
and/or it can be dependent with the Lagrangian equations. This
means, the identities are possible among Lagrangian equations and
the completion functions:
\begin{equation}\label{GI}
\Gamma_\alpha^i(\phi)\partial_iS(\phi)+\Gamma_\alpha^a(\phi)\tau_a(\phi)\equiv
0 \, .
\end{equation}
The coefficients $\Gamma_\alpha^i(\phi), \Gamma_\alpha^a(\phi)$ are
the matrices of the differential operators. We interpret the above
relations as the most general form of Noether identities in the
system where the Lagrangian equations are incomplete in the above
mentioned sense. In Section 4 we provide explicit examples of the
field theories where the gauge identities involve the completion
functions.

It may seem that the equations (\ref{tau-a}) could be considered on equal footing with the Lagrangian equations
\begin{equation}\label{LE}
    \partial_i S(\phi)\approx 0 \, ,
\end{equation}
so we have just general non-Lagrangian theory with equations of motion (\ref{tau-a}) and (\ref{LE}) with gauge identities (\ref{GI}).
General gauge algebra for not necessarily Lagrangian theory defined just by equations of motion is well known \cite{Kazinski:2005eb}.
The issues of locality of gauge algebra of not necessarily Lagrangian theories are described in details in reference
\cite{Kaparulin:2011xy}.
In fact, there is a subtlety which makes a difference between the general non-Lagrangian equations and incomplete Lagrangian system.
In the latter case, the mass shell is defined by Lagrangian equations (\ref{LE}), while the completion equations (\ref{tau-a})
do not restrict the solutions of Lagrangian equations, given the boundary conditions.
Once the mass shell is a zero locus of critical points of the action $S(\phi)$,
the gauge symmetry should leave the action invariant, while the symmetry of the general equations is defined irrespectively
to the existence of action at all. It is the difference which leads to the constraints on gauge parameters as we shall see below.

Notice that the gauge identity generators $\Gamma$ are defined by
the relations (\ref{GI}) modulo  natural ambiguity. The
generators $\Gamma$ and $\Gamma'$ are considered equivalent once
they differ by certain on-shell vanishing terms:
\begin{eqnarray}
  \Gamma'^i_\alpha(\phi)-\Gamma^i_\alpha(\phi)&=&  E^{ij}_\alpha
(\phi)\partial_iS (\phi)+  E^{ia}_\alpha (\phi)\tau_a(\phi)\, , \quad  E^{ij}_\alpha = -E^{ji}_\alpha;  \label{GG`1}\\
 \Gamma'^a_\alpha(\phi)-\Gamma^a_\alpha(\phi)&=&  E^{ab}_\alpha
(\phi)\tau_b(\phi) -  E^{ia}_\alpha(\phi) \partial_iS (\phi) \, ,
\quad  E^{ab}_\alpha = -E^{ba}_\alpha. \label{GG`2}
\end{eqnarray}
If the identity  generators $\Gamma$ are replaced by $\Gamma'$ in
relations (\ref{GI}), all the coefficients $E$ will drop out from
the identities. To put it different, the right hand sides of the
relations (\ref{GG`1}), (\ref{GG`2}) are understood as trivial
generators of gauge identities. If the Lagrange equations were
complete in the above mentioned sense, no completion functions
$\tau_a$ would be admitted by the theory. If $\tau_a$ were not
involved, the relations (\ref{GG`1}) would correspond to the usual
definition of trivial gauge generators in Lagrangian theory. In the
incomplete case, the gauge identities can involve completion
functions, that is why the definition is modified of the trivial
generators.

Also notice that the change of the generating set of completion functions (\ref{tplusds})
results in the corresponding change of the gauge generators $\Gamma^i_\alpha$
\begin{equation}\label{GdSchange}
    \tau_a(\phi)\, \mapsto \, \tau'_a(\phi)=\tau_a(\phi)+\theta_a^i(\phi)\partial_iS(\phi) \, , \quad \Gamma^i_\alpha\,\mapsto\, \Gamma'^i_\alpha=\Gamma^i_\alpha+\theta_a^i(\phi)\Gamma^a_\alpha
\end{equation}

The left hand sides of the Lagrangian equations $\partial_iS(\phi)$
and completion functions $\tau_a(\phi)$ constitute the generating
set of the on-shell vanishing quantities, as stated by relations
(\ref{Completion}). Much like that, the operators
$\Gamma^i_\alpha,\Gamma^a_\alpha$ (\ref{GI}) are assumed to form the
generating set for the gauge identities. Any generator of gauge
identity is assumed to be a linear combination of $\Gamma_\alpha$
modulo trivial generators:
\begin{equation}\label{G-complete}
    L^i(\phi)\partial_iS(\phi)+L^a(\phi)\tau_a(\phi)\equiv 0\quad
    \Rightarrow\quad L^i(\phi)\approx
    k^\alpha(\phi)\Gamma^i_\alpha(\phi),\,\,\,L^a(\phi)\approx
    k^\alpha(\phi)\Gamma^a_\alpha(\phi)\,.
\end{equation}
In this article we assume that the generators $\Gamma_\alpha$ are
independent in the sense that they can not be linearly combined with
on-shell non vanishing coefficients into a trivial generator. In the
other words, the gauge identities (\ref{GI}) are assumed
irreducible. In principle, this assumption restricts generality,
though no example is known at the moment of field theory with
reducible unfree gauge symmetry.

Now, let us see that the gauge identities (\ref{GI}) define the
gauge symmetry of the action, while the gauge transformation
parameters cannot be free, once the completion functions are
involved. Consider the gauge transformation of the fields
\begin{equation}\label{GT}
    \delta_\epsilon\phi^i=\Gamma^i_\alpha(\phi)\epsilon^\alpha \, ,
\end{equation}
where $\epsilon^\alpha$ are the gauge transformation parameters.
With the account of the identity (\ref{GI}), the gauge variation of
the action reads:
\begin{equation}\label{GTS}
    \delta_\epsilon S(\phi)\equiv\epsilon^\alpha\Gamma^i_\alpha(\phi)\partial_iS(\phi) \equiv-\epsilon^\alpha\Gamma^a_\alpha(\phi)\tau_a(\phi)\,
    .
\end{equation}
If the completion functions $\tau_a(\phi)$ were not not involved
into the gauge identities (\ref{GI}), i.e. if $\Gamma^a_\alpha=0$,
the action would remain intact under the gauge variation (\ref{GT})
with free parameters $\epsilon^\alpha$. Once $\Gamma^a_\alpha\neq
0$, the action is invariant under the transformation (\ref{GT}) if
the gauge parameters are constrained by the equations:
\begin{equation}\label{eps-constr}
\epsilon^\alpha\Gamma^a_\alpha(\phi)=0.
\end{equation}
As we see, the gauge identities (\ref{GI}) involving completion functions (\ref{tau-a})
 result in the gauge symmetry of the action
\begin{equation}\label{S-inv}
     \delta_\epsilon
S(\phi)\equiv 0 \, ,
\end{equation}
though the gauge parameters have to be constrained by equations
(\ref{eps-constr}). The quantity $\Gamma^a_\alpha$, being involved
in the gauge identity (\ref{GI}) as a coefficient at completion
function, defines the restriction imposed onto gauge parameter. With
this regard, we term $\Gamma^a_\alpha$ as operator of gauge parameter constraint.

As we have mentioned above, the generating set of completion functions is defined modulo lhs of Lagrangian equations (\ref{tplusds}).
This leads to the ambiguity in the definition of the gauge symmetry generators (\ref{GdSchange}).
This ambiguity does not contribute to the
gauge transformations because the gauge parameters are unfree (\ref{eps-constr}).

As we have already said,  the completion equations (\ref{tau-a}) do not impose
 restrictions on the solutions of the Lagrange equations (\ref{LE}) with given boundary conditions.
 This means that the mass shell remains invariant under the transformations which leave the action intact.
 In particular, any on shell vanishing local quantity should remain
  vanishing on shell after the gauge transformation
 (\ref{GT}),(\ref{eps-constr}),
 \begin{equation}\label{shell-inv}
    O(\phi)\approx 0 \quad\Rightarrow\quad \delta_\epsilon
    O(\phi)\approx 0 \, .
\end{equation}
Once the unfree gauge variation (\ref{GT}), (\ref{eps-constr})
vanishes on shell of the local quantity $O$, off shell this means
\begin{equation}\label{Off-sh}
  \delta_\epsilon O(\phi)\approx 0
   \quad\Leftrightarrow\quad\Gamma^i_\alpha\partial_iO(\phi)=V^i_\alpha(\phi)\partial_iS(\phi)
   +V^a_\alpha(\phi)\tau_a(\phi)+ W_a(\phi)\Gamma^a_\alpha(\phi) \,.
\end{equation}
For the gauge symmetry with constrained gauge parameters, the last two
terms can arise once the gauge symmetry is unfree. This is a distinction from the theory with unconstrained gauge parameters.
Also notice that the  last one of these two terms does not vanish on shell.

For the requirement of the gauge invariance of the mass shell
(\ref{shell-inv}) to be satisfied, it is sufficient that it is
satisfied for the generating set of local quantities vanishing on
the shell. The set includes Lagrangian equations $\partial_i
S(\phi)$ and the completion functions (\ref{tau-a}), so these
quantities have to be on shell gauge invariant,
\begin{equation}\label{eq-inv}
    \delta_\epsilon\partial_iS(\phi)\approx 0,
    \quad\delta_\epsilon\tau_a(\phi)\approx 0 \, .
\end{equation}
Making use of relations (\ref{Off-sh}) we get the off shell
action of gauge generators on Lagrangian equations and completion
functions
\begin{equation}\label{Gtau}
\Gamma^i_\alpha(\phi)\partial_i \tau_a(\phi)= R_{\alpha
a}^i(\phi)\partial_iS(\phi)+R_{\alpha a}^b(\phi)\tau_b(\phi)+
W_{ab}(\phi)\Gamma^b_\alpha (\phi)
 \, .
\end{equation}
\begin{equation}\label{GdS}
\Gamma^i_\alpha(\phi)\partial_i\partial_jS(\phi)= R_{\alpha
j}^i(\phi)\partial_iS(\phi)+R_{\alpha j}^b(\phi)\tau_b(\phi)+
W_{jb}(\phi)\Gamma^b_\alpha (\phi) \, .
\end{equation}
The specifics of the unfree gauge symmetry is seen at first in the
terms which involve the operator of gauge parameter constraint
$\Gamma^a_\alpha$. These terms do not necessarily vanish on shell.  They
originate from the fact the the mass shell is invariant under the
transformations with the parameters restricted by the equations
(\ref{eps-constr}). We also mention that the structure functions
$W_{ab}(\phi)$ involved in the off shell non-vanishing  terms in
relation (\ref{Gtau}) are on shell symmetric
\begin{equation}\label{W-symm}
    W_{ab}(\phi) - W_{ba}(\phi)\approx 0 \, .
\end{equation}
This property can be deduced by making use of consequences of the identity (\ref{GI})
and the assumption of completeness of the generating set (\ref{G-complete}).

Let us also notice, that the structure functions $U$ in the right
hand sides of the relations (\ref{Gtau}), (\ref{GdS}) vanish in the
linear theory. Unlike that, the structure functions $W$ (that do not
have any analogue in the theory with free gauge parameters) do not
necessarily vanish even in the linear theory. In Section 4 we
provide an explicit example of linear field theory with the
structure relations (\ref{Gtau}) involving non-trivial structure
functions $W$.

Notice that any gauge symmetry transformation of the mass shell
should be spanned by the gauge generators $\Gamma^i_\alpha$ once the
gauge parameters are restricted by the equations (\ref{eps-constr}).
This fact is a consequence of the completeness assumption
(\ref{G-complete}). For the gauge symmetry transformations
(\ref{GT}), this means the commutator of the generators should be
spanned by the generators modulo trivial ones defined by relations
(\ref{GG`1}), and up to the terms that do not contribute to the
transformations once the parameters are unfree (\ref{eps-constr}).
Explicitly, the commutators read
\begin{eqnarray}
\nonumber \Gamma^i_\alpha(\phi)\partial_i\Gamma^j_\beta(\phi)-\Gamma^i_\beta(\phi)\partial_i\Gamma^j_\alpha(\phi)&=&
    U_{\alpha\beta}^\gamma(\phi)\Gamma^j_\gamma(\phi)\\
     + E_{\alpha\beta}^{aj}(\phi)\tau_a(\phi) + E_{\alpha\beta}^{ij}(\phi)
    \partial_iS(\phi) &+& R_{\alpha a}^{j}(\phi)\Gamma^a_\beta(\phi) - R_{\beta a}^{j}(\phi)\Gamma^a_\alpha(\phi) \,. \label{GG}
\end{eqnarray}
The following relations should be hold between the unfree gauge
symmetry generators $\Gamma^i_\alpha$ and the gauge parameter
constraints operators $\Gamma^a_\alpha$:
\begin{eqnarray}\label{GiGa}
\nonumber  \Gamma^i_\alpha(\phi)\partial_i\Gamma_\beta^a(\phi)-\Gamma^i_\beta(\phi)\partial_i\Gamma_\alpha^a(\phi)  &=&
 U_{\alpha\beta}^\gamma(\phi)\Gamma_\gamma^a(\phi) + \\
  R_{\alpha}{}_ b^a(\phi)\Gamma^b_\beta(\phi) - R_{\beta}{ }^a_b(\phi)\Gamma^b_\alpha(\phi)&+&
  E_{\alpha\beta}^{ab}(\phi)\tau_b(\phi) - E_{\alpha\beta}^{ai}(\phi)\partial_iS(\phi) \,.
\end{eqnarray}
To get this relation, we contract the gauge identity (\ref{GI}) with
a test function with the property (\ref{eps-constr}), and then
compute the gauge variation of obtained expression. After that the
roles of the test function and gauge parameter are interchanged. The
difference of the gauge variations is an identity between the
Lagrangian equations and completion functions (\ref{tau-a}). Under
the assumption (\ref{G-complete}) of completeness this means that
the coefficients at completion functions are proportional to the
gauge generators $\Gamma_\alpha^a$ modulo trivial generators
(\ref{GG`1}) and gauge parameter constraints (\ref{eps-constr}). The
formula (\ref{GiGa}) express no more than this fact.

In the first relations of the unfree gauge algebra (\ref{GI}), two
extra constituents are involved -- the completion functions
(\ref{tau-a}), and the operators of gauge parameter constraints
(\ref{eps-constr}). These quantities have no direct analogue either
in Lagrangian gauge theory with unconstrained gauge parameter or in
the general non-Lagrangian gauge system. Also notice that
non-Lagrangian field equations could be also incomplete in the same
sense, in principle. Hence the unfree gauge symmetry could occur for
non-variational field equations, though no explicit examples are
known yet of this phenomenon. Further compatibility conditions are
possible for the unfree gauge algebra  relations involving higher
structure functions, much like the case with free gauge
transformations parameters.

With unconstrained gauge parameters, all the structure relations of
gauge algebra are generated by the BV master equation. In
non-Lagrangian case, instead of the master action, the operator $Q$
is constructed of the BRST transformation \cite{Kazinski:2005eb}
with the initial data defined by the classical field equations,
their gauge symmetries and gauge identities. The relation $Q^2=0$
replaces the master equation in this case, and all the gauge algebra
structure relations are generated by this equation. The set of
ghosts is more general in non-Lagrangian theory than in the
Lagrangian case. For the recipe of BRST embedding for not
necessarily Lagrangian field equations we refer to the article
\cite{Kazinski:2005eb}. In particular, anti-fields are assigned to
the field equations, not to the fields, while there is no pairing
between fields and equations once the equations are not supposed to
be the variational derivatives of any action. As the gauge
identities are not necessarily paired with gauge symmetries in
non-Lagrangian theory, the ghosts are not necessarily dual to the
ghost anti-fields in this case \cite{Kazinski:2005eb}. Much like the
non-Lagrangian field theory, the case of unfree gauge symmetry would
require a more general set of ghosts than the case without
constraints on the gauge parameters. In this article, we do not work
out a procedure for the field-anti-field BV-BRST embedding of a
system with general unfree gauge symmetry. We restrict the
consideration by the case where no higher structure functions
appear, and the FP recipe is sufficient. This is briefly considered
in the next section.

\section{Faddeev-Popov path integral for systems  with unfree gauge symmetry}
We  begin with a geometric remark that the ghosts are the
coordinates on the fibers of the same bundle as the gauge
transformation parameters \cite{Kazinski:2005eb}. The difference is
that the Grassmann parity of the ghosts is shifted by 1 with respect
to the parity of the gauge parameters, and the ghosts are assigned
with the ghost number grading 1. Because of this geometric reason,
the ghosts have to be constrained by the same equations as the gauge
transformation parameters (\ref{eps-constr}). So, once the ghosts
$C^\alpha$ are assigned to the unfree gauge transformations
(\ref{GT}), they should be subjected to the equations
\begin{equation}\label{C-constr}
\Gamma^a_\alpha(\phi)C^\alpha=0 \, , \qquad gh(C^\alpha)=1\, , \quad \varepsilon(C^\alpha)= 1\, ,
\end{equation}
where $\Gamma^a_\alpha(\phi)$ is the operator of gauge parameter constraint (\ref{eps-constr}).
The path integration should be done over the surface of the ghost constraints, not by free ghosts $C^\alpha$.

Now, let us discuss the gauge fixing in the theory with unfree gauge
symmetry. Let us denote the number of unfree gauge parameters by
$m$, and the number of gauge parameter constraints will be  $n$. If
one could locally solve the equations (\ref{eps-constr}) and find
the unconstrained gauge parameters, without symmetries for
symmetries, their number would be $\bar{m}=m-n$. This means,
$\bar{m}$ independent conditions are required to fix the gauge.
Denote the independent gauges $\chi^I(\phi)$. The index $I$ is
condensed, so it includes the space coordinates $x^\mu$. The
dimension of digital part of the index should be $\bar{m}$. If we
use the independent gauge fixing conditions,
 the number of unfree gauge parameters will exceed the number number of gauges, so FP matrix will be rectangular,
\begin{equation}\label{FPM}
   \frac{\delta_\epsilon\chi^I}{\delta\epsilon^\alpha} =\Gamma^i_\alpha(\phi)\partial_i\chi^I (\phi)\,
    .
\end{equation}
The rectangular matrix cannot be invertible, nor can it have the determinant, while it can be considered non-degenerate in certain sense.
Consider the equation for the null-vectors $u_I(\phi)$ of the matrix,
\begin{equation}\label{FPN}
    \Gamma^i_\alpha(\phi)\partial_i\chi^I u _I(\phi)\approx 0\,
    .
\end{equation}
If the general solution for $u_I$ does not involve arbitrary
functions of all the space-time coordinates $x^\mu$, the FP matrix
is considered non-degenerate. To put it different, the general gauge
orbit is transverse to zero locus of $\chi^I(\phi)$. Some gauge
variations (\ref{GT}) can be tangential to the surface
$\chi^I(\phi)=0$, though the corresponding parameters are
constrained much stronger that just by the condition
(\ref{eps-constr}), so they can involve arbitrary functions of less
than $d$ coordinates in $d$-dimensional space-time. This definition
of admissible gauge fixing condition is applicable also in the case
with free gauge parameters. For example, the Lorentz gauge in
Maxwell electrodynamics has the d'Alembert operator as the FP
matrix. Relation (\ref{FPN}) in this case is just d'Alembert
equation, so the general solution involves arbitrary functions of
$d-1$ coordinates (e.g., Cauchy data).

Given the admissible gauge fixing conditions, the anti-ghosts
\begin{equation}\label{barC}
    \bar{C}_I\,, \qquad gh(\bar{C}_I)=-1,\quad \varepsilon(\bar{C}_I)-1
\end{equation}
 are assigned to $\chi^I(\phi)$.
 The number of the anti-ghosts is $\bar{m}$, and it is less than the number of ghosts, $\bar{m}=m-n$,
where $m$ is the number of ghosts and $n$ is the number of ghost constraints (\ref{C-constr}).
Unlike the  case of unconstrained gauge symmetry, there is no pairing between ghosts and anti-ghosts
if the gauges are chosen independent.

Given the gauges, and FP matrix, we can consider the adjustment of the FP path integral to the case of unfree gauge symmetry.
Once the ghosts are subject to the equations (\ref{C-constr}) in the case of unfree gauge symmetry,
the FP path integral for the transition amplitude has to be restricted to the ghost constraint surface, so it reads
\begin{equation}\label{FP-int}
    Z_{FP}= \int \prod_{i,\alpha, a , I}[d\phi^i dC^\alpha d \bar{C}_I ] \, \,
    \delta(\chi^I(\phi))\delta(\Gamma_\beta^a(\phi)C^\beta)\, \,
\exp{\frac{i}{\hbar}\{  S(\phi) +  \bar{C}_I\Gamma^i_\alpha\partial_i\chi^I(\phi)C^\alpha
    \}} \, .
\end{equation}
This amplitude  could be viewed as an implicit expression of the
standard FP integral over the original fields and independent ghosts
introduced for the gauge transformations with free parameters if one
could find the local unconstrained parametrization of the gauge
symmetry. If, for example, one could explicitly rearrange the gauge
generators into unconstrained ones and zero operators, the delta
functions of ghost constraints would just remove the ghosts for the
vanishing parameters, so the expression (\ref{FP-int}) would
reproduce the usual FP amplitude.

Consider Fourier representation for the delta-functions of the gauges  and ghost constraints
\begin{eqnarray}
\prod_{I}\delta(\chi^I(\phi)) &=& \int \prod_{I}[d\pi_I] \exp{\frac{i}{\hbar}\pi_J \chi^J(\phi)}\,, \quad \varepsilon (\pi_I)=gh(\pi_I)=0 ;  \label{delta-chi}\\
\prod_a\delta(\Gamma_\beta^a(\phi)C^\beta)&=&\int\prod_{a}[d\bar{C}_a]
\exp{\frac{i}{\hbar}\bar{C}_b\Gamma^b_\beta(\phi)C^\beta} \,
, \quad \varepsilon(\bar{C}_a)=1, \,\, gh(\bar{C}_a)=-1\,
. \label{delta-gammaC}
\end{eqnarray}
Substituting (\ref{delta-chi}), (\ref{delta-gammaC}) into (\ref{FP-int}), we bring the FP integral to Feynman's form
\begin{equation}\label{ZFP}
Z= \int [d\varphi]\exp{\frac{i}{\hbar}S_{FP}(\varphi)} \, , \qquad\varphi=(\phi^i, \pi_\alpha, C^\alpha,\bar{C}_I, \bar{C}_a) \, ,
\end{equation}
where the FP action reads
\begin{equation}\label{FP-act}
    S_{FP}= S(\phi) + \pi_I\chi^I(\phi)+
    \bar{C}_I\Gamma^i_\alpha(\phi)\partial_i\chi^I(\phi)C^\alpha +
    \bar{C}_a\Gamma^a_\alpha (\phi) C^\alpha \, .
\end{equation}
Once the Fourier multipliers $\bar{C}_a$ to the ghost constraints
$\Gamma^a_\alpha(\phi)C^\alpha$ have the ghost number $-1$, these
can be considered as anti-ghosts, on equal footing with the
anti-ghosts $\bar{C}_I$ assigned to the gauge fixing conditions
$\chi^I(\phi)$. With this regard, the total number of anti-ghosts in
the FP action (\ref{FP-act}) becomes equal to the total number of
ghosts, while the ghosts are not constrained anymore. The matrix of
ghost-anti-ghost bilinear form in the FP action (\ref{FP-act}) is
squared, and it is non-degenerate. So, the integral (\ref{ZFP}) is
regular both in ghosts and zero ghost number variables, including
original fields and Lagrange multipliers to the gauges.

Let us discuss the independence of the FP path integral of the
choice of gauge fixing conditions $\chi^I(\phi)$. The BRST symmetry
is an appropriate tool for the control of gauge independence of the
path integral, so let us seek for the   BRST transformation.

Given the gauge identities (\ref{GI}), the natural candidate for the BRST symmetry generator read
\begin{equation}\label{QFP}
    Q= C^\alpha\Gamma^i_\alpha\partial_i +
    \pi_I\frac{\partial}{\partial\bar{C}_I}+\tau_a\frac{\partial}{\partial\bar{C}_a}
    + o(C^2) \, .
\end{equation}
Let us consider for simplicity the abelian case, when the generators
$\Gamma^i_\alpha$ commute, and  do not act on $\Gamma^a_\alpha$, and
therefore no $C^2$ terms can appear in the BRST transformation. In
this case the FP action is
 obviously $Q$-invariant, because of the gauge identity (\ref{GI}),
\begin{equation}\label{Q-abS}
    Q S_{FP}=C^\alpha\left( \Gamma^i_\alpha\partial_i S
    +\Gamma^a_\alpha\tau_a\right) + \bar{C}_IC^\alpha C^\beta\left(\Gamma^i_\alpha\Gamma^j_\beta\partial_i\partial_j\chi^I \right)\equiv
    0\,.
\end{equation}
$Q$ does not square to zero identically unless the structure
functions $W_{ab}$ vanish in relation (\ref{Gtau}):
\begin{equation}\label{Q-ab2}
    Q^2=C^\alpha \Gamma^i_\alpha\partial_i\tau_a \frac{\partial}{\partial\bar{C}_a}\equiv
    W_{ab}\Gamma^b_\alpha C^\alpha
    \frac{\partial}{\partial\bar{C}_a} \,.
\end{equation}
Notice that the path integral (\ref{FP-int}) is localized at the
ghost constraint surface (\ref{C-constr}), where the BRST
transformation (\ref{QFP}) is truly nilpotent. If  we considered the
constraints on the ghosts (\ref{C-constr}) as a part of mass shell,
the BRST-generator $Q$ would square to zero on shell. In the
theories with open gauge algebra, the BRST symmetry of the gauge
fixed theory typically holds only on shell.

The BRST symmetry of the FP action (\ref{FP-act}) means that the
path integral is independent from the choice of gauge fixing
condition $\chi(\phi)$. This can be seen in the same way as for the
case with unconstrained gauge transformation parameters. Consider
the infinitesimal change of the gauge
\begin{equation}\label{dchi}
    \chi(\phi)\mapsto\chi(\phi)+\delta\chi(\phi) \, .
\end{equation}
Let us make the BRST transformation of all the fields, including
ghosts, anti-ghosts and Lagrange multipliers $\varphi=(\phi^i,\pi_I,
C^\alpha,\bar{C}_I,\bar{C}_a)$ with the transformation parameter
$\delta\Psi$ induced by the change of gauge (\ref{dchi}):
\begin{equation}\label{delta-psi}
    \varphi\mapsto\varphi_\Psi = \varphi +\delta\varphi, \quad \delta\varphi= (Q\varphi)\delta\Psi\,, \qquad \delta\Psi= \frac{i}{\hbar}\bar{C}_I\delta\chi^I(\phi) \, .
\end{equation}
Once the FP action (\ref{FP-act}) is BRST-invariant (\ref{QFP}), the infinitesimal change of fields can contribute
to the path integral only through the transformation Jacobian.
Up to the first order in $\delta\chi(\phi)$, the Jacobian reads
\begin{equation}\label{Jac}
    \det{\left(\frac{\partial\varphi_\Psi}{\partial\varphi}\right)}
    =\exp\frac{i}{\hbar}\left(\pi_I\delta\chi^I(\phi)+ \bar{C}_I \Gamma^i_\alpha (\phi)
    \frac{\partial \delta\chi^I}{\partial\phi^i} +
    (\textrm{div} Q)\bar{C}_I\delta\chi^I\right) \, ,
\end{equation}
where $\textrm{div} Q$ is a divergence of the BRST transformation
vector  $Q$ (\ref{QFP}). The divergence of $Q$ is a simplest
characteristic class of any gauge system \cite{Lyakhovich:2004kr}.
Complete classification of characteristic classes of gauge systems
can be found in \cite{Lyakhovich:2009qq}. The divergence of BRST
transformation is usually termed as a modular class. The one-loop
anomaly is known to be proportional to the modular class. If the
modular class vanishes (hence, the theory is free from anomaly), the
Jacobian (\ref{Jac}) reproduces the change of the gauge fixing
condition (\ref{dchi}) in the FP path integral (\ref{FP-int}). In
this way, one can see that the FP path integral (\ref{FP-int}) with
the gauge $\chi(\phi)$ is connected by the change of fields
(\ref{delta-psi}) with the integral involving the gauge
$\chi(\phi)+\delta\chi(\phi)$.

As we have seen, the FP path integral (\ref{FP-int}) does not depend
on the choice of gauge fixing condition in the sense that the
infinitesimal change of the gauge can be compensated by the change
of the integration variables. The path integral quantization recipe
(\ref{FP-int}) applies to the theories with unfree gauge symmetry
when no higher structure functions are involved in the gauge
algebra. Once the higher structure functions are involved, e.g.,
when the gauge transformations do not commute off shell, the
extension of the BV field-anti-field formalism has to be worked out
for the case of unfree gauge algebra. This  will be done elsewhere,
while some clues to the extension are discussed in the conclusion of
the article.

\section{Examples}
In this section we exemplify the general structures of theories with
unfree gauge symmetry by two linear models: traceless spin two free
field with the action proposed in \cite{Alvarez:2006uu}, and the
Maxwell-like Lagrangian \cite{Campoleoni:2012th} for the tracefull
second rank tensor field. Even in the linear models,  the
distinctive structures of the unfree gauge symmetry turn out
non-trivial. At first, we demonstrate the non-trivial completion
functions (\ref{tau-a}) in these models. Then, we observe that the
gauge identities (\ref{GI}) involve, besides the Lagrangian
equations and gauge symmetry generators, also completion functions
and the operators of gauge symmetry constraints (\ref{eps-constr}).
We also see that the structure functions $W_{ab}$ (\ref{Gtau}),
(\ref{W-symm}) can be non-trivial already at linear level in the
models with unfree gauge symmetry.

The considered models admit at least three different ways of
quantization that allows one to verify the results by the cross
check. First, besides the unfree irreducible parametrization of
gauge symmetry with the parameters constrained by the differential
equations, these models admit an alternative parametrization with
free gauge parameters, though with the symmetry of symmetry. The FP
quantization rules involving ghosts-for-ghosts are well known for
the theories with reducible gauge symmetry. Second, by inclusion
appropriate auxiliary fields into the action, both models can be
equivalently reformulated in the way with unconstrained irreducible
gauge symmetry. In this form, the usual FP quantization rules apply,
while the auxiliary fields can be eliminated from the path integral
by imposing the gauge fixing conditions such that the extra fields
are forced to vanish. And third, the modification of FP recipe
(\ref{ZFP}), (\ref{FP-act}) can be directly applied to both models
 in the original form with unfree gauge symmetry, without any
reformulation. As we shall see, all three ways lead to the same
result in these models, so the examples confirm proposed ansatz
(\ref{ZFP}), (\ref{FP-act}). Even though the models are linear, and
the ghost contributions to the path integral reduce to the
determinants of the field independent operators, it can be
considered as a reasonable test for the correctness of the path
integral (\ref{ZFP}), (\ref{FP-act}) for the theories with unfree
gauge symmetry. The reason is that the perturbative inclusion of
interactions would deform the ghost terms of free theory, not
replace them by the structures with a different constant part.

For the second example -- the Maxwell-like Lagrangian -- besides the
free theory we consider the specific cubic vertex found in the
article \cite{Francia:2016weg}. A complete classification of
consistent cubic interactions of higher spin fields is established
in the article \cite{Metsaev:2007rn} making use of light-cone
formalism. The new vertex does not correspond to any cubic
interaction in this classification, though it seems admissible from
the viewpoint of the usual Noether procedure of inclusion
interactions. While the vertex is local, it can be removed by a
nonlocal change of fields noticed in the article
\cite{Francia:2016weg}.
 Given the general BRST differential (\ref{QFP}) of the theory
with the unfree gauge symmetry, we shall demonstrate that the above
mentioned vertex is BRST exact, i.e. it is the BRST variation of
local quantity. This means, the vertex is trivial from the viewpoint
of the local gauge field theory that explains why it does not have
the place on the list of admissible interactions of the article
\cite{Metsaev:2007rn}.

\subsection{Linearized unimodular gravity}
Consider symmetric traceless second rank tensor field
$h_{\mu\nu}(x)$, $h^\nu{}_{\nu}(x)\equiv0$ in $d=4$ Minkowski space
with the action
\begin{equation}\label{S-uni}
    S[h(x)]=\int L d^4x,\qquad L=\frac{1}{2}(\partial_\mu
    h_{\nu\rho}\partial^\mu
    h^{\nu\rho}-2\partial_{\mu}h_{\nu\rho}\partial^{\nu}h^{\mu\rho})\,.
\end{equation}
The signature of the metric is mostly negative. The Lagrange
equations  read
\begin{equation}\label{LE-uni}
    \frac{\delta S}{\delta h^{\mu\nu}}=-\Box h_{\mu\nu}+\partial_{\mu}\partial^\rho h_{\rho\nu}
    +\partial_{\nu}\partial^\rho h_{\rho\mu}-\frac{1}{2}\eta_{\mu\nu}\partial^\rho\partial^\lambda h_{\rho\lambda}\approx
    0\,, \quad \Box= \partial_\rho\partial^\rho\, .
\end{equation}
Taking the divergence of the equations, we arrive at the
differential consequence (Cf. (\ref{ddd})):
\begin{equation}\label{LE-uni-ddd}
    \partial_\mu\tau\approx 0\,,\qquad
    \tau=\partial^\nu\partial^\lambda
    h_{\nu\lambda}\,.
\end{equation}
Provided for zero boundary conditions for the fields at the space
infinity, we obtain a single completion function
\begin{equation}\label{tau-uni}
    \tau (x)\approx0\,,
\end{equation}
which has the form (\ref{tau-a}) where the condensed index $a$ is
just the Minkowski space point $x$. Any on shell vanishing local
quantity is spanned by the Lagrangian equations (\ref{LE-uni}) and
completion function (\ref{tau-uni}):
\begin{equation}\label{mass-shell-uni}
   O(h,\partial h, \partial^2 h, \partial^3 h, \ldots)\approx 0 \quad\Leftrightarrow\quad
    O= \hat{V}{}^{\mu\nu}\frac{\delta S}{\delta
    h^{\mu\nu}} + \hat{V}\tau\, ,
\end{equation}
where $\hat{V}{}^{\mu\nu}, \hat{V} $ are the local differential
operators. This means that completion functions and left hand sides
of Lagrangian equations constitute the generating set for the
on-shell vanishing local quantities.

The generatings (\ref{LE-uni}), (\ref{tau-uni}) of this set are
dependent. The gauge identities (\ref{GI}) between the Lagrangian
equations (\ref{LE-uni}) and completion function (\ref{tau-uni})
read
\begin{equation}\label{GI-uni}
    \partial^\nu\frac{\delta S}{\delta h^{\mu\nu}}-\frac{1}{2}\partial_\mu\tau\equiv0\,.
\end{equation}
From this relation one can find the identity generators  (\ref{GI})
in this model
\begin{equation}\label{G-uni-ident}
    \Gamma^{i}_\rho\equiv\Gamma^{\mu\nu}_\rho=\delta^\mu{}_\rho\partial^\nu+\delta^\nu{}_\rho\partial^\mu\,,\qquad
    \Gamma^a_\rho\equiv\Gamma_\rho=\partial_\rho\,.
\end{equation}
In accordance with the general structure of unfree gauge algebra
described in Section 2, the coefficient at the completion function
in the gauge identity (\ref{GI}) should define the operator of the
gauge parameter constraint (\ref{eps-constr}), while the coefficient
at the Lagrangian equation defines the generator of unfree gauge
symmetry (\ref{GT}). Given the explicit form of the identity
generators (\ref{G-uni-ident}), the unfree gauge transformation
(\ref{GT}) and the gauge parameter constraint (\ref{eps-constr}) in
this model should read
\begin{equation}\label{gt-uni-con}
    \delta_\epsilon
    h_{\mu\nu}=\partial_\mu\epsilon_{\nu}+\partial_\mu\epsilon_{\mu}\,,\qquad
    \partial_\mu\epsilon^\mu=0\,.
\end{equation}
In this way we see that the involvement of the completion function
in the gauge identity (\ref{GI-uni}) defines the linearized $T$-diff
as a gauge symmetry of the model. The unfree gauge variation
(\ref{gt-uni-con}) obviously leaves the
 action
 (\ref{S-uni}) intact
\begin{equation}\label{gt-S-uni}
    \delta_\epsilon S[h(x)]=\int \tau \partial_\mu \epsilon^\mu=0\,.
\end{equation}
Let us also verify the gauge-invariance of the mass shell and
identify the structure function $W$ (\ref{Gtau}) of this model
\begin{equation}\label{mass-shell-inv-uni}
    \delta_\epsilon \left(\frac{\delta S}{\delta
    h^{\mu\nu}}\right)=\frac{1}{2}\partial_\mu\partial_\nu(\partial_\rho\epsilon^\rho)=0\,,\qquad\delta_\epsilon
    \tau=2\Box\partial_\nu\epsilon^\nu=0\,.
\end{equation}
As is seen, the structure function (\ref{Gtau}) does not vanish,
\begin{equation}\label{W-simm-uni}
    W=2\Box\,.
\end{equation}
Once the d'Alembertian is self-adjoint, the structure function $W$
is symmetric indeed, cf. (\ref{W-symm}).

\vspace{0.2 cm}

Now, consider the path integral quantization of the model
(\ref{S-uni}). At first, we shall apply the recipe (\ref{ZFP}),
(\ref{FP-act}) to get the transition amplitude for the theory
proceeding from the original action (\ref{S-uni}) and unfree gauge
symmetry (\ref{gt-uni-con}). Then, we shall consider the two
equivalent reformulations of the model. One of these reformulations
makes use the same action (\ref{S-uni}) while the gauge symmetry is
parameterized in a different way. The gauge parameters are
unconstrained but the symmetry is reducible:  it admits a sequence
of gauge symmetry for symmetry. This allows one to quantize the
model along the usual lines, by introducing ghosts for ghosts. One
more reformulation makes use of the fact that the model
(\ref{S-uni}) can be viewed as a partially gauge-fixed version of
linearized Einstein's gravity with the partial gauge $h^\mu_\mu=0$.
Choosing the complete gauge fixing conditions involving this partial
gauge, one can explicitly integrate out the trace of $h_{\mu\nu}$
and get the path integral for the model (\ref{S-uni}).

To simplify the comparison of the results of three methods, it is
convenient to impose the same gauge fixing conditions in the sector
of the original fields in all the schemes. We choose the independent
gauge-fixing conditions,
\begin{equation}\label{chi}
    \chi{}^i=\partial_jh^{ji}-\frac{1}{2}\partial^ih{}^{j}{}_{j}=0\,
    .
\end{equation}
Here, and below in this section, the Latin indices $i,j=1,2,3$ label
the space components of the Minkowski tensors or coordinates.

Let us begin with applying the quantization receipt (\ref{ZFP}),
(\ref{FP-act}) to the model (\ref{S-uni}). The unfree gauge symmetry
generators (\ref{GT}) and gauge parameter constraint operators
(\ref{eps-constr})  are defined for this model by relations
(\ref{gt-uni-con}). Substituting (\ref{S-uni}), (\ref{gt-uni-con}),
and the gauge-fixing conditions (\ref{chi}) into the general
prescription (\ref{ZFP}), (\ref{FP-act}) we arrive at the path
integral
\begin{equation}\label{FP-uni}\begin{array}{l}\displaystyle
    Z=\int[dhdb^idC_\mu d\bar{C}^i d\bar{C}] \exp \frac{i}{\hbar}S_{FP} \,,\\[4mm]\displaystyle
    S_{FP}=\int\bigg(L+\bar{C}{}^i\Delta
    C_{i}+\bar{C}\partial_\mu C^\mu+(\partial^j
    h_{ij}-1/2\partial_ih^j{}_j)b^i\bigg)d^4x\,,\qquad
    \Delta=\sum_{i=1}^3\partial_i^2\,.
\end{array}\end{equation}
The FP action (\ref{FP-uni}) is invariant with respect to the action
of the BRST symmetry operator (\ref{QFP}),
\begin{equation}\label{Q-uni}
    Q=(\partial^\mu C^\nu+\partial^\nu C^\mu)\frac{\delta}{\delta h^{\mu\nu}}+b{}^i\frac{\delta\,}{\delta \bar{C}{}^i}+
    \tau\frac{\delta}{\delta \bar{C}}\,.
\end{equation}

Now, let us consider an alternative parametrization of gauge
symmetry. The action (\ref{S-uni}) admits an unconstrained reducible
parametrization of gauge symmetry
\begin{equation}\label{gt-uni}
    \delta_\xi h_{\mu\nu}=\partial_{\mu}\partial^\rho\widetilde{\xi}_{\rho\nu}+
    \partial_{\nu}\partial^\rho\widetilde{\xi}_{\rho\mu}\,,\qquad\widetilde{\xi}_{\mu\nu}=
    \frac{1}{2}\varepsilon_{\mu\nu\rho\sigma}\xi^{\rho\sigma}\,, \qquad \delta_\xi S(h)\equiv0\, , \quad\forall\xi.
\end{equation}
The transformation parameter is an antisymmetric tensor
$\xi_{\mu\nu}=-\xi_{\mu\nu}$ whose components are arbitrary
functions of space-time coordinates. The gauge symmetry
(\ref{gt-uni}) is reducible. The gauge transformations for the gauge
parameters read
\begin{equation}\label{gt-uni-1}
    \delta_{\xi{}^{(1)}}\xi_{\mu\nu}=\partial_\mu\xi{}^{(1)}{}_\nu-\partial_\nu\xi{}^{(1)}{}_\mu\,,\qquad
    \delta_{\xi{}^{(2)}}\xi{}^{(1)}{}_\mu=\partial_\mu\xi{}^{(2)}\,.
\end{equation}
Once the gauge transformations are reducible, the minimal set of the
BV fields and anti-fields includes ghosts for ghosts and conjugate
anti-fields. The gradings read of the fields and anti-fields
\begin{equation}\label{gh-n-uni}
    \text{gh}\,h_{\mu\nu}=0\,,\qquad
    \text{gh}\,C_{\mu\nu}=1\,,\qquad
    \text{gh}\,C^{(1)}{}_{\mu}=2\,,\qquad \text{gh}\,C^{(2)}=3\,.
\end{equation}
\begin{equation}\label{gh-a-uni}
    \text{gh}\,h^\ast{}_{\mu\nu}=-1\,,\qquad
    \text{gh}\,C^\ast{}_{\mu\nu}=-2\,,\qquad
    \text{gh}\,C^{(1)\ast}{}_{\mu}=-3\,,\qquad
    \text{gh}\,C^{(2)\ast}=-4\,.
\end{equation}
The minimal BV-action for the theory reads
\begin{equation}\label{SBV-min-uni}
    S_{min}=\int\bigg(L+{h}^\ast{}_{\mu\nu}(\partial^{\mu}\partial_\rho\widetilde{C}^{\rho\nu}+
    \partial^{\nu}\partial_\rho\widetilde{C}^{\rho\mu})+C^\ast{}_{\mu\nu}(\partial^\mu C^{(1)}{}^\nu-
    \partial^\nu C^{(1)}{}^\mu)+C^{(1)\ast}{}_{\mu}\partial^\mu
    C^{(2)}\bigg)d^4x.
\end{equation}
To introduce the fields and anti-fields in non-minimal sector, we
assume that all the gauges for original fields and ghosts are
independent. This is a slight deviation from the usual scheme of BV
quantization of the theories with reducible gauge symmetry where the
reducibility of gauge fixing conditions are assumed to follow the
reducibility pattern of gauge symmetry generators. In fact, this
assumption can be bypassed, as we see in this example. The
independent gauges can exist both for the fields and for the ghosts,
even if the gauge symmetries, and symmetries for symmetries are
reducible. In the case at hands, it is convenient to use the
independent gauges because this simplifies comparison with the
result obtained in terms of irreducible generators of gauge
symmetry.

Let us choose the same independent gauge fixing conditions
(\ref{chi}) for the original fields $h_{\mu\nu}$ as we applied
above. For the ghosts $C_{\mu\nu}$ and the next level ghosts for
ghosts $C_\mu^{(1)}$, we can also choose the irreducible gauge
fixing conditions:
\begin{equation}\label{gauges-uni-red}
    \chi{}^{(1)}{}_i(C)\equiv C_{0i}=0\, ,\quad
    \chi{}^{(2)}(C^{(1)})\equiv C^{(1)}_0=0 \, .
\end{equation}
 Now let us introduce anti-ghosts
$\bar{C}$ and Lagrange multipliers $b$ for every gauge fixing
condition. The ghost number of the multiplier should  be opposite to
the number of the gauge fixing condition, while the ghost number of
the anti-ghost is shifted by $-1$ with respect to the number of the
multiplier. The anti-field has the opposite number to that of the
field shifted by minus one. All that means, we introduce the
following variables of the non-minimal sector:
\begin{equation}\label{non-min-gh-uni}
    \text{gh}\,\bar{C}_{i}=-1\,,\quad
    \text{gh}\,\bar{C}^{(1)}{}_{i}=-2\,,\quad
    \text{gh}\,\bar{C}^{(2)}=-3\,,\quad
    \text{gh}\,b_i=0\,,\quad
    \text{gh}\,b^{(1)}=-1\,,\quad
    \text{gh}\,b^{(2)}=-2\,.
\end{equation}
\begin{equation}\label{non-min-a-uni}
    \text{gh}\,\bar{C}^\ast{}_{i}=0\,,\quad
    \text{gh}\,\bar{C}^{(1)\ast}{}_{i}=1\,,\quad
    \text{gh}\,\bar{C}^{(2)\ast}=2\,,\quad
    \text{gh}\,b^\ast{}_i=-1\,,\quad
    \text{gh}\,b^{(1)\ast}=0\,,\quad
    \text{gh}\,b^{(2)\ast}=1\,.
\end{equation}
As all the gauges are independent, there is no gauge symmetry for
the anti-ghosts, and the Lagrange multipliers for the gauges. That
is why no ghosts for ghosts are introduced in the non-minimal
sector.

The non-minimal BV-action is introduced in the form
\begin{equation}\label{SBV-nm-uni}
    S_{nonmin}=S_{min}+\int\bigg(\bar{C}{}^{\ast i}b{}_i+\bar{C}{}^{(1)}{}^{\ast i}{b}{}^{(1)}{}_i+
    \bar{C}{}^{(2)\ast}{b}{}^{(2)}\bigg)d^4x.
\end{equation}
Given the gauge fixing conditions (\ref{gauges-uni-red}), the
gauge-fixing fermion reads
\begin{equation}\label{psi-uni}
    \psi=\int\bigg(\bar{C}{}^{i}(\partial^j h_{ij}-1/2\partial_ih^j{}_j)+\bar{C}{}^{(1)}{}^{i}C_{0i}+
    \bar{C}{}^{(2)}C^{(1)}{}_0\bigg)d^4x.
\end{equation}
The gauge-fixing for anti-fields, being defined as
$\varphi^*=\partial\psi/\partial \varphi$, reads:
\begin{equation}\label{a-phi-uni}\begin{array}{c}\displaystyle
    h{}^\ast{}_{0i}=0\,\qquad h{}^\ast{}_{ij}=-1/2(\partial_i\bar{C}_j+\partial_j\bar{C}_i-\delta_{ij}\partial_k\bar{C}^{k})\,;
    \\[3mm]\displaystyle
    C{}^\ast{}_{ij}=0\,,\qquad C{}^\ast{}_{0i}=\bar{C}{}^{(1)}{}_i\,,\qquad
    C{}^{(1)\ast}{}_i=0\,,\qquad
    C{}^{(1)\ast}{}_0=\bar{C}{}^{(2)}\,;\\[3mm]\displaystyle
    \bar{C}{}^\ast{}_{i}=\partial^j
    h_{ij}-1/2\partial_ih^j{}_j\,,\qquad
    \bar{C}{}^{(1)\ast}{}_{i}=C_{0i}\,,\qquad
    \bar{C}{}^{(2)\ast}=C{}^{(1)}{}_0\,.
\end{array}\end{equation}
Gauge fixed action reads
\begin{equation}\label{S-psi-uni}\begin{array}{c}\displaystyle
    S_{\psi}=\int\bigg(L-(\partial_i\bar{C}_j+\partial_j\bar{C}_i-\delta_{ij}\partial_k\bar{C}{}^{k})
    \partial^{i}\partial_\rho\widetilde{C}^{\rho j}+\bar{C}{}_{0i}(\partial^0 C^{(1)}{}^i-
    \partial^i C^{(1)}{}^0)+\bar{C}{}^{(2)}\partial_0
    C^{(2)}+\\[3mm]\displaystyle
    +
    (\partial^j h_{ij}-1/2\partial_ih^j{}_j)b{}^{i}+C_{0i}b{}^{(1)}{}^{i}+
    C^{(1)}{}_0b{}^{(2)}\bigg)d^4x.
\end{array}\end{equation}
Simplifying this expression, we get
\begin{equation}\label{S-psi-simpl}\begin{array}{c}\displaystyle
    S_{\psi}=\int\bigg(L+1/2\varepsilon^{ikl}\bar{C}_i\partial_0\Delta C_{jk}+\bar{C}{}_{0i}\partial^0 C^{(1)}{}^i+\bar{C}{}^{(2)}\partial_0
    C^{(2)}+\\[3mm]\displaystyle
    +
    (\partial^j h_{ij}-1/2\partial_ih^j{}_j)b{}^{i}+C_{0i}b{}^{(1)}{}^{i}+
    C^{(1)}{}_0b{}^{(2)}\bigg)d^4x.
\end{array}\end{equation}
Let us show the path integral with the action above,
\begin{equation}\label{Z-psi-uni}
    Z=\int[d\varphi]\exp\frac{i}{\hbar}S_\psi(\varphi)\,,\qquad
    \varphi=(h{}_{\mu\nu}\,, C_{\mu\nu}\,,\bar{C}{}_i\,,b{}^i\,, C{}^{(1)}{}_\mu\,,
    \bar{C}{}^{(1)}{}_i\,, b^{(1)i}\,,
    C{}^{(2)}\,,\bar{C}{}^{(2)}\,, b^{(2)})\,,
\end{equation}
can brought to the form (\ref{FP-uni}). The main steps are as
follows. First, the time derivatives at ghosts $C{}_{ij}$ and
$C{}^{(1)i}$ are absorbed by the field redefinition,
\begin{equation}\label{gh-change-1}
    \mathcal{C}{}_{ij}=\partial_0C_{ij}\,,\qquad
    \mathcal{C}^{(1)}=\partial_0C^{(1)i}\,.
\end{equation}
The differential change of variables should change the integration
measure by the Jacobian, being the determinant of corresponding
differential operator. As we use the same operator for changing the
variables of the opposite Grassmann parity, the Jacobians cancel
each other.  After that, the reducibility ghosts of minimal and
non-minimal sector (except $C{}^{(2)}$ and $\bar{C}{}^{(2)}$) are
integrated out making use of corresponding gauge fixing conditions.
This does not add any factor to the integration measure at this
step. The intermediate result for the path integral reads
\begin{equation}\label{Step2}\begin{array}{c}\displaystyle
    Z=\int[d\varphi]\exp\bigg\{\frac{i}{\hbar}\int\bigg(L+1/2\varepsilon^{ikl}\bar{C}_i\Delta \mathcal{C}_{kl}+\bar{C}{}^{(2)}
    \partial_0C^{(2)}+(\partial^j
    h_{ij}-1/2\partial_ih^j{}_j)b{}^{i}\bigg)d^4x\bigg\}\,,\\[4mm]\displaystyle
    \varphi=(h{}_{\mu\nu}\,,b^i\,,\mathcal{C}{}_{ij}\,,\bar{C}{}^{i}\,,C{}^{(2)}\,,\bar{C}{}^{(2)})\,.
\end{array}\end{equation}
At the third step, the path integral by the ghost number $3$, $-3$
variables $C{}^{(2)}$, $\bar{C}{}^{(2)}$ is replaced by the
equivalent expression involving the new ghost $1$, $-1$ variables
$\mathcal{C}^0\,,\bar{\mathcal{C}}$,
\begin{equation}\label{Step3}\begin{array}{c}\displaystyle
    \int[d\bar{C}{}^{(2)}dC{}^{(2)}]\exp\bigg\{\frac{i}{\hbar}\int\bar{C}{}^{(2)}
    \partial_0C^{(2)}d^4x\bigg\}=\int[d\overline{\mathcal{C}} d\mathcal{C}^0]\exp\bigg\{\frac{i}{\hbar}\int
    \bar{\mathcal{C}}\partial_0\mathcal{C}^0d^4x\bigg\}\,,\\[5mm]\displaystyle
    \text{gh}\,\mathcal{C}^0=1\,,\qquad \text{gh}\,\bar{\mathcal{C}}=-1.
\end{array}\end{equation} At the fourth step, we perform the off-diagonal shift of the
ghost field $\mathcal{C}^0$,
\begin{equation}\label{C2-change}
\mathcal{C}{}^{0}\qquad\rightarrow\qquad
\mathcal{C}{}^{0}+1/2\partial{}_0{}^{-1}\varepsilon^{ijk}\partial{}_i\mathcal{C}_{jk}\,.
\end{equation}
The Jacobian of this variable change is unit, so the integration
measure is preserved. After the transformations (\ref{Step3}),
(\ref{C2-change}) are done, we rewrite the path integral
(\ref{Step2}) in the form
\begin{equation}\label{Step34}\begin{array}{c}\displaystyle
    Z=\int[d\varphi]\exp\bigg\{\frac{i}{\hbar}\int\bigg(L+1/2\varepsilon^{ikl}\bar{C}_i\Delta \mathcal{C}_{kl}+\bar{\mathcal{C}}
    (\partial_0\mathcal{C}^{0}+1/2\varepsilon^{ijk}\partial{}_i\mathcal{C}_{jk})+\\[4mm]\displaystyle+(\partial^j
    h_{ij}-1/2\partial_ih^j{}_j)b{}^{i}\bigg)d^4x\bigg\}\,,\qquad
    \varphi=(h{}_{\mu\nu}\,,b^i\,,\mathcal{C}{}_{ij}\,,\bar{C}{}^{i}\,,\mathcal{C}{}^{0}\,,\bar{\mathcal{C}})\,.
\end{array}\end{equation}
The modified FP action (\ref{FP-uni}) follows from this formula if
the ghosts $\mathcal{C}^0\,,1/2\varepsilon^{ijk}\mathcal{C}_{jk}$
are considered as the time and space components of 4-vector
$C^{\mu}=(\mathcal{C}^0,1/2\varepsilon^{ijk}\mathcal{C}_{jk})$.

Consider one more way to deduce the path integral for the theory
(\ref{S-uni}). Introduce the action functional for the linearized
Einstein's gravity,
\begin{equation}\label{S-E}
    S_{gr}[h(x)]=\int L_{gr} d^4x\,,\qquad L_{gr}=\frac{1}{2}(\partial_\mu
    h_{\nu\rho}\partial^\mu
    h^{\nu\rho}-2\partial_{\mu}h_{\nu\rho}\partial^{\nu}h^{\mu\rho}+2\partial_\mu
    h^{\mu\nu}\partial^\nu h-\partial_\mu h\partial^\nu
    h)\,.
\end{equation}
The dynamical field in the model is the tracefull second-rank
symmetric tensor $h_{\mu\nu}(x)$, and the notation is used $h\equiv
h^\mu{}_\mu\,.$ The action functional (\ref{S-E}) is invariant under
linearized diffeomorphism gauge transformation,
\begin{equation}\label{}
    \delta_\xi h_{\mu\nu}=\partial_\mu\xi_\nu+\partial_\nu\xi_\mu\,.
\end{equation}
The gauge parameter is the vector $\xi$, which is unconstrained. The
convenient gauge fixing for the linearized Einstein's theory
includes relations (\ref{chi}) and zero trace condition,
\begin{equation}\label{g-fix-ein}
    \chi_i\equiv\partial^jh_{ij}-\frac12\partial_ih^j{}_j=0\,,\qquad
    h=0\,.
\end{equation}
With this gauge imposed, the trace of the metric is excluded, and
the classical theory coincides with the model (\ref{S-uni}) in the
gauge (\ref{chi}). Consider now the conventional FP for the model
(\ref{S-E}) in the gauge (\ref{g-fix-ein}):
\begin{equation}\label{FP-gr}
    S_{FP}=\int\bigg(L_{gr}+\bar{C}{}^i\Delta
    C_{i}+\bar{C}\partial_\mu C^\mu+(\partial^j
    h_{ij}-1/2\partial_ih^j{}_j)b^i+h\,b\bigg)d^4x\,.
\end{equation}
The multiplier $b$ and the trace of the metric $h$ can be integrated
out. After that, the path integral (\ref{FP-gr}) takes the form
(\ref{FP-uni}). In this way, one can see that the conventional FP
path integral for linearized gravity in the gauge $h^\mu_\mu = 0$
and (\ref{chi}) reproduces the answer (\ref{FP-uni}) constructed by
the general recipe (\ref{ZFP}), (\ref{FP-act}) for the linearized
unimodular gravity. This example confirms once again the general
prescription (\ref{ZFP}), (\ref{FP-act}) for path integral in the
theory with unfree gauge symmetry.

\subsection{Maxwell-like theory of symmetric tensor field.} Consider the symmetric tracefull second rank
tensor field $h_{\mu\nu}(x)$, $h^\mu{}_\mu=h$ in $d=4$ Minkowski
space. The action reads as in the previous case (\ref{S-uni}), where
the tensor $h_{\mu\nu}$ is tracefull. Many of the relations of the
previous subsection hold true for this model if the tensor is
understood as tracefull. So, we provide the relations which cannot
be obtained in this way, otherwise we refer to the previous section.

The Lagrangian equations for the Maxweel-like model of second rank
tensor field read
\begin{equation}\label{LE-MaxLike}
    \frac{\delta S}{\delta h^{\mu\nu}}\equiv-\Box h_{\mu\nu}+\partial_{\mu}\partial^\rho
    h_{\rho\nu}+\partial_{\nu}\partial^\rho
    h_{\rho\mu}\approx0\,.
\end{equation}
These equations have differential consequence (\ref{LE-uni-ddd}).
The completion function is the same as in the previous example,
$\tau=\partial_\mu\partial_\nu h^{\mu\nu}$ (\ref{tau-uni}). The
gauge identities (\ref{GI}) have slightly different form,
\begin{equation}\label{GI-ML}
    \partial^\nu\frac{\delta S}{\delta h^{\mu\nu}}-\partial_\mu\tau\equiv0\,.
\end{equation}
The identity generators read
\begin{equation}\label{}
    \Gamma^{\mu\nu}_\rho=\delta^\mu{}_\rho\partial^\nu+\delta^\nu{}_\rho\partial^\mu\,,\qquad
    \Gamma_\rho=2\partial_\rho\,.
\end{equation}
There is additional factor 2  in $\Gamma_\rho$ comparing to
(\ref{GI-uni}), (\ref{G-uni-ident}). The constrained gauge
transformation is the linearized T-diff (\ref{gt-uni-con}). The set
of on-shell vanishing quantities includes the Lagrangian equations
(\ref{LE-MaxLike}) and completion functions (\ref{tau-uni}), which
are gauge-invariant.

The quantization of the Maxwell-like theory proceeds along the same
lines as the linearized unimodular gravity. All the quantization
schemes turn out equivalent in this case much like the previous one,
so we write down the FP action (\ref{FP-act}) in the gauge
(\ref{chi}) omitting the details of derivation
\begin{equation}\label{FP-MaxLike}
    S_{FP}=\int\bigg(L+\bar{C}{}^i\Delta
    C_{i}+\bar{C}\partial_\mu C^\mu+(\partial^j
    h_{ij}-1/2\partial_ih^j{}_j)b^i\bigg)d^4x\,.
\end{equation}
The BRST symmetry generator (\ref{QFP})  reads in this case
\begin{equation}\label{Q-MaxLike}
    Q=(\partial^\mu C^\nu+\partial^\nu C^\mu)\frac{\delta}{\delta h^{\mu\nu}}+b{}^i\frac{\delta\,}{\delta \bar{C}{}^i}+
    2\tau\frac{\delta}{\delta
    \bar{C}}\,.
\end{equation}
The action (\ref{FP-MaxLike}) is BRST invariant with respect to this
transformation.

Let us consider the Maxwell-like Lagrangian with specific cubic
vertex found in the article \cite{Francia:2016weg}:
\begin{equation}\label{S3}
    L(g)=\frac{1}{2}(\partial_\mu
    h_{\nu\rho}\partial^\mu
    h^{\nu\rho}-2\partial_{\mu}h_{\nu\rho}\partial^{\nu}h^{\mu\rho})-g(\partial_\mu\partial_\nu
    h^{\mu\nu})h_{\rho\lambda}h^{\rho\lambda}\,,
\end{equation}
where $g$ the coupling constant.
 Up to the first order in $g$, the
unfree gauge transformation for the action read,
\begin{equation}\label{gt-3}
    \delta_\epsilon
    h_{\mu\nu}=\partial_\mu\epsilon_\nu+\partial_\nu\epsilon_\mu\,,\qquad
    \partial_\mu\epsilon^\mu+g(\partial_\mu\epsilon_\nu+\partial_\nu\epsilon_\mu)h^{\mu\nu}=0\,.
\end{equation}
Notice that the symmetry transformation remains unchanged at this
level, while the gauge parameter constraint operator is deformed.
The vertex (\ref{S3}) is gauge invariant with respect to the gauge
transformation above, with account for the deformation of the
constraint. This cubic interaction seems admissible and non-trivial
from the viewpoint of the Noether procedure for inclusion of gauge
invariant interaction applied in the work \cite{Francia:2016weg}
along the usual lines of theory with the free gauge parameters, and
accompanied by deformation of the gauge parameter constraint. On the
other hand, this vertex  does not fit into the known classification
of cubic interactions of higher spin fields \cite{Metsaev:2007rn}.
In the article \cite{Francia:2016weg} this discrepancy is explained
in the following way. The authors observe that this interaction
vertex can be interpreted as due to non-local redefinitions of the
fields,
\begin{equation}\label{non-loc-h}
    h_{\mu\nu}\quad\rightarrow\quad
    h_{\mu\nu}+g\frac{\partial_\mu\partial_\nu}{\Box}\,h_{\rho\lambda}h^{\rho\lambda}\,.
\end{equation}
Proceeding from this nonlocal substitution, it is concluded that
inclusion of the vertex does not alter the physical properties of
the model. Below, we shall demonstrate, without any recourse to
non-local manipulations, that the local vertex (\ref{S3}) is trivial
indeed, given the BRST symmetry (\ref{Q-MaxLike}) of the FP action
(\ref{FP-MaxLike}).

At first, let us construct the FP action (\ref{FP-act}) for the
model with the interaction vertex (\ref{S3}). To do that, we have to
upload into the general formula (\ref{FP-act}) all the specific
ingredients of the model: the Lagrangian (\ref{S3}); the gauge
fixing conditions (\ref{chi}); the gauge generators and gauge
parameter constraint operators (\ref{gt-3}). The result reads
\begin{equation}\label{FP-MaxLike-3}\begin{array}{l}\displaystyle
    S_{FP}(g)\equiv
    S^{(0)}_{FP}+gS^{(1)}_{FP}=\phantom{\int\bigg(}\\[5mm]\displaystyle\int\bigg(L(g)+\bar{C}{}^i\Delta
    C_{i}+\bar{C}(\partial_\mu C^\mu+g(\partial_\mu C_{\nu}+\partial_\nu C_\mu)h^{\mu\nu}+(\partial^j
    h_{ij}-1/2\partial_ih^j{}_j)b^i\bigg)d^4x\,.
\end{array}\end{equation}
The cubic terms include both the original vertex and the ghost
contributions:
\begin{equation}\label{FP-MaxLike-33}
    S{}^{(1)}_{FP}=\int \bigg(-(\partial_{\mu}{\partial_\nu})h^{\mu\nu}h_{\rho\lambda}h^{\rho\lambda}+\bar{C}(\partial_\mu C_{\nu}+\partial_\nu
    C_\mu)h^{\mu\nu}\bigg)d^4x\,.
\end{equation}
Even though the symmetry is abelian, and the gauge generators are
constants, the action involves the cubic term with ghosts
$\bar{C}(\partial_\mu C_{\nu}+\partial_\nu
    C_\mu)h^{\mu\nu}$. This term originates from the contribution of the
    gauge parameter constraint operator to the FP action (\ref{FP-act}). As
    the unfree gauge parameter is constrained at interacting level by the equation
    (\ref{gt-3}) involving $h$, the field contributes to the ghost
    term. It is easy to see that the full cubic part of the FP
    action (\ref{FP-MaxLike-33}) is BRST-exact with respect to the
    BRST  differential of the free theory (\ref{Q-MaxLike}):
\begin{equation}\label{FP-cubic-triv}
    S^{(1)}_{FP}=Q\Psi\, , \qquad\Psi =-\frac{1}{2}\int \bar{C}\, h_{\rho\lambda}h^{\rho\lambda}\, d^4x\,.
\end{equation}
The potential $\Psi$ is local, so the vertex is trivial indeed from
the viewpoint of local BRST cohomology.

\section{Concluding remarks}
Let us first summarize the results of the article, and then discuss the remaining problems.

In this article we study the general phenomenon of the gauge
symmetry with unfree gauge parameters. We proceed from the
conjecture that the system of Lagrangian equations is incomplete in
certain sense: given the boundary conditions, the local on shell
vanishing quantities exist such that they do not reduce to
combination of the equations and their derivatives. We choose the
generating set for on-shell vanishing local quantities
(\ref{Completion}) which includes Lagrangian equations and
completion functions (\ref{tau-a}). The latter quantities vanish on
shell, while they are not differential consequences of Lagrangian
equations. In general, the gauge identities of the theory involve
both Lagrangian equations and completion functions (\ref{GI}). It is
the structure of gauge identities which leads to the constraints
(\ref{eps-constr}) on the gauge symmetry parameters of the action
(\ref{GTS}). In the usual case, when the Lagrangian field theory
does not admit the completion functions, the gauge algebra involves
the two primary constituents: the action functional and the
generating set of the gauge symmetry transformations. Given these
two constituents, all the higher structures of gauge algebra are
defined by the compatibility conditions of gauge identities. Once
the Lagrangian equations admit completion functions, the gauge
identities (\ref{GI}) involve two more ingredients: the completion
functions (\ref{tau-a}), (\ref{Completion}) and the gauge parameter
constraint operators (\ref{eps-constr}). With this regard, the
higher structure relations of the unfree generated gauge algebra
involve more structure functions comparing to the algebra with
unconstrained gauge parameters. We deduce the structure relations of
the unfree gauge algebra up to the level which corresponds to the
Lie algebra in the case of unconstrained gauge symmetry. We observe
that the unfree gauge algebra can involve non-trivial structure
constants which do not necessarily vanish even in the linear theory.
This does not have a direct analogue even at the linear level of the
theories with unconstrained gauge parameters. As an example, we can
mention the structure function $W_{ab}$ involved in the relation
(\ref{Gtau}) which follows from the fact that mass shell is
invariant under the transformations with unfree gauge parameters. If
all the structure functions are constants, we suggest the extension
of the FP path integral construction to the case of unfree gauge
symmetry. The path integral (\ref{ZFP}), (\ref{FP-act}) explicitly
involves the operators of gauge parameter constraint. The FP action
is BRST invariant, while the BRST differential (\ref{QFP}) has a
distinction from the case of unconstrained gauge symmetry, as the
completion functions are explicitly involved. The BRST invariance
ensures the gauge independence of the path integral.

In Section 4, we consider two examples of the field theories with
unfree gauge transformation parameters. Both models admit
alternative parametrization of gauge symmetry with unconstrained
parameters, though with the gauge symmetry of symmetry. Also, by
inclusion of auxiliary fields, they can be equivalently reformulated
as theories with  irreducible gauge symmetry and unconstrained gauge
parameters. In this way, one can verify the path integral
quantization recipe (\ref{ZFP}), (\ref{FP-act}) by comparing the
transition amplitude with the ones deduced by usual FP rules based
on the alternative gauge symmetry parameterizations. All the answers
coincide for the amplitude. Notice that the second example also
demonstrates how the BRST symmetry of the theory with unfree gauge
parameters can be helpful for separation of nontrivial interaction
vertices from the trivial ones. The vertex has been previously known
in this model which looks eligible from the viewpoint of Noether
procedure, while it should not appear from the view point of known
classification of admissible cubic interactions. The BRST complex,
which makes a due account for the gauge parameter constraints and
completion functions, identifies this vertex as BRST exact, and
thereby trivial.

Let us mention some remaining problems concerning general structure of the field theories with unfree gauge parameters,
and possible solutions.

At first, notice that the quantization recipe (\ref{ZFP}),
(\ref{FP-act}) involves independent gauge fixing conditions. In many
cases the independent gauge fixing conditions cannot be consistent
with Poincar\'e or AdS symmetry of the theory with unfree gauge
symmetry. The reason is obvious: for example, given the vector gauge
parameter in $d$ dimensions restricted by the transversality
equation, the number of independent gauge fixing conditions should
be $d-1$, so they cannot be tensors. It is unlikely to find $d-1$
appropriate scalars to fix the gauge. Explicitly covariant gauges
are admissible, for example $\partial \cdot h$ in Maxwell-like
theory \cite{Campoleoni:2012th} or in the model of traceless higher
spin fields \cite{Skvortsov:2007kz},
 These gauge fixing conditions are obviously over-complete and therefore they should be on-shell reducible.
 The reducibility is obvious indeed, $\partial \cdot \partial \cdot h\approx 0$. Reducibility of the gauge condition would require
 the ghosts for ghosts with the higher negative ghost numbers (anti-ghosts),
 while no ghosts for ghosts are introduced with positive ghost numbers.
 This asymmetry between ghost and anti-ghost sectors does not have direct counterpart in the case of reducible gauge
 symmetry where the over-complete set of gauge generators is mirrored by the reducible set of gauge fixing conditions.
 That is why the ghosts-for-ghosts are accompanied by anti-ghosts for anti-ghosts.
 It should be examined, however, that the asymmetry between ghost and anti-ghost sector is consistent with the usual physical interpretation of
  BRST cohomology groups.

 The second open problem is the extension of the BV field-anti-field formalism
 to the class of field theories with unfree gauge symmetry.
To begin with the problem, the field-anti-field content of the theory has to be modified comparing
 to the case of gauge symmetry with unconstrained parameters.
If the field-anti-field space remained the same, the BV master
equation would generate the usual gauge algebra relations where no
place is left for the operators of gauge parameter constraints
(\ref{eps-constr}). The general idea of finding an appropriate
field-anti-field space extension is that the ghost constraints
(\ref{C-constr}) and completion equations (\ref{tau-a}) have to be
considered as equations of motion, on equal footing with the
original Lagrangian equations. This brings the theory, at least for
a while, to the realm of not necessarily Lagrangian systems. In not
necessarily Lagrangian case, the anti-fields are assigned to the
equations \cite{Kazinski:2005eb}, not to the fields, while in
Lagrangian case this would the same. The specifics of the ghost
constraint equation (\ref{C-constr}) is that it has the ghost number
one. This means, the corresponding anti-field should have zero ghost
number, while the anti-field to the completion equation would have
the ghost number $-1$. These two extra-anti-fields are the dual
variables, and they have the opposite parity, so it is natural to
expect that they should be taken as conjugate with respect to the
anti-bracket. Proceeding from this general setup, we expect to
develop the BV formalism for the systems with general unfree gauge
symmetry in the future work.

\vspace{0.1cm} \noindent \textbf{Acknowledgements}. We thank
V.~Abakumova, M.~Grigoriev, D.~Francia, K.~Mkrtchan, and A.~Sharapov
for discussions on various issues related to this work. The work is
partially supported by Tomsk State University Competitiveness
Improvement Program. The work of SLL is supported by the project
3.5204.2017/6.7 of Russian Ministry of Science and Education.

\end{document}